\input{aipcheck}

\documentclass[
    ,final            
  ]
  {aipproc}

\layoutstyle{8x11single}

\begin{document}

\title{On the splitting of nucleon effective masses at high isospin density:
 reaction observables}

\classification { 21.30.Fe, 21.65.+f, 24.10.Jv, 25.75.-q }
\keywords      {Asymmetric nuclear matter, neutron/proton effective
mass splitting, Lane potential, relativistic heavy ion collisions, particle 
emission, collective flows}

\author{M.Di Toro}{
  address={Laboratori Nazionali del Sud INFN, via S.Sofia 62, I-95123
Catania, Italy
and Dipartimento di Fisica e Astronomia, Universita' di Catania}
}

\author{M.Colonna}{
  address={Laboratori Nazionali del Sud INFN, via S.Sofia 62, I-95123
Catania, Italy
and Dipartimento di Fisica e Astronomia, Universita' di Catania}
}

\author{J.Rizzo}{
  address={Laboratori Nazionali del Sud INFN, via S.Sofia 62, I-95123
Catania, Italy
and Dipartimento di Fisica e Astronomia, Universita' di Catania}
}

\begin{abstract}
We review the present status of the nucleon effective mass splitting
$puzzle$ in asymmetric matter, with controversial predictions within both
non-relativistic $and$ relativistic approaches to the effective in medium 
interactions. Based on microscopic transport simulations we suggest
some rather sensitive observables in collisions of asymmetric (unstable)
ions at intermediate ($RIA$) energies: i) Energy systematics of Lane 
Potentials; ii) Isospin content of fast emitted nucleons; iii) Differential
Collective Flows. Similar measurements for light isobars (like $^3H-^3He$)
could be also important.

\end{abstract}

\maketitle


\section{Isospin Momentum dependence in Skyrme forces}

We start from a discussion on the isospin dependence of these
widely used non-relativistic effective interactions \cite{VautherinPRC3}.
In a Skyrme-like parametrization  the symmetry
term has the form:
\begin{equation}
\epsilon_{sym} \equiv {E_{sym} \over A} (\rho) = {\epsilon_F(\rho) \over 3} 
+ {C(\rho) \over 2} ~ {\rho \over \rho_0}
\label{kipotsky}
\end{equation}
with the function $C(\rho)$, in the potential part, given by:
\begin{eqnarray}
{C(\rho) \over {\rho_0}} = - {1 \over 4} \Big[t_0 (1 + 2 x_0) + {t_3 \over 6} 
(1 + 2 x_3)~ \rho^\alpha \Big] \nonumber\\
 + {1 \over {12}}
\Big[t_2 (4 + 5 x_2) - 3 t_1 x_1 \Big] \Big( {{3 \pi^2} \over 2} \Big)
^{2/3}~ \rho^{2/3} \equiv {1 \over \rho_0} \Big[ C_{Loc}(\rho) + 
C_{NLoc}(\rho) \Big]
\label{sky}
\end{eqnarray}
with $\alpha > 0$ and the usual Skyrme parameters.
We remark that the second term is related to isospin effects on the momentum 
dependence \cite{LyonNPA627}. 

\begin{figure}[htb]
\begin{minipage}[t]{67mm}
\includegraphics*[scale=0.40]{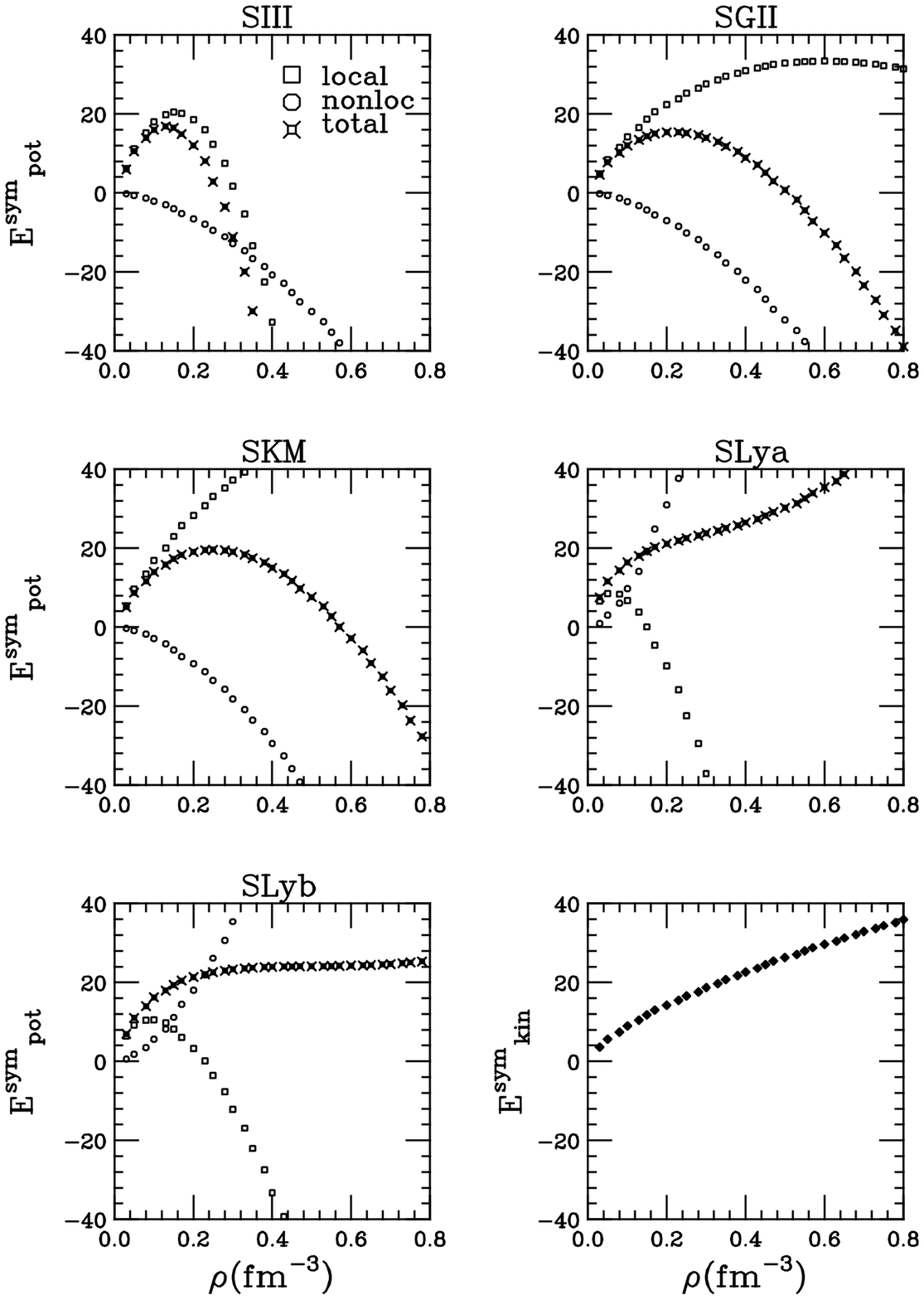}
\end{minipage}
\hspace{\fill}
\begin{minipage}[t]{67mm}
\includegraphics*[scale=0.40]{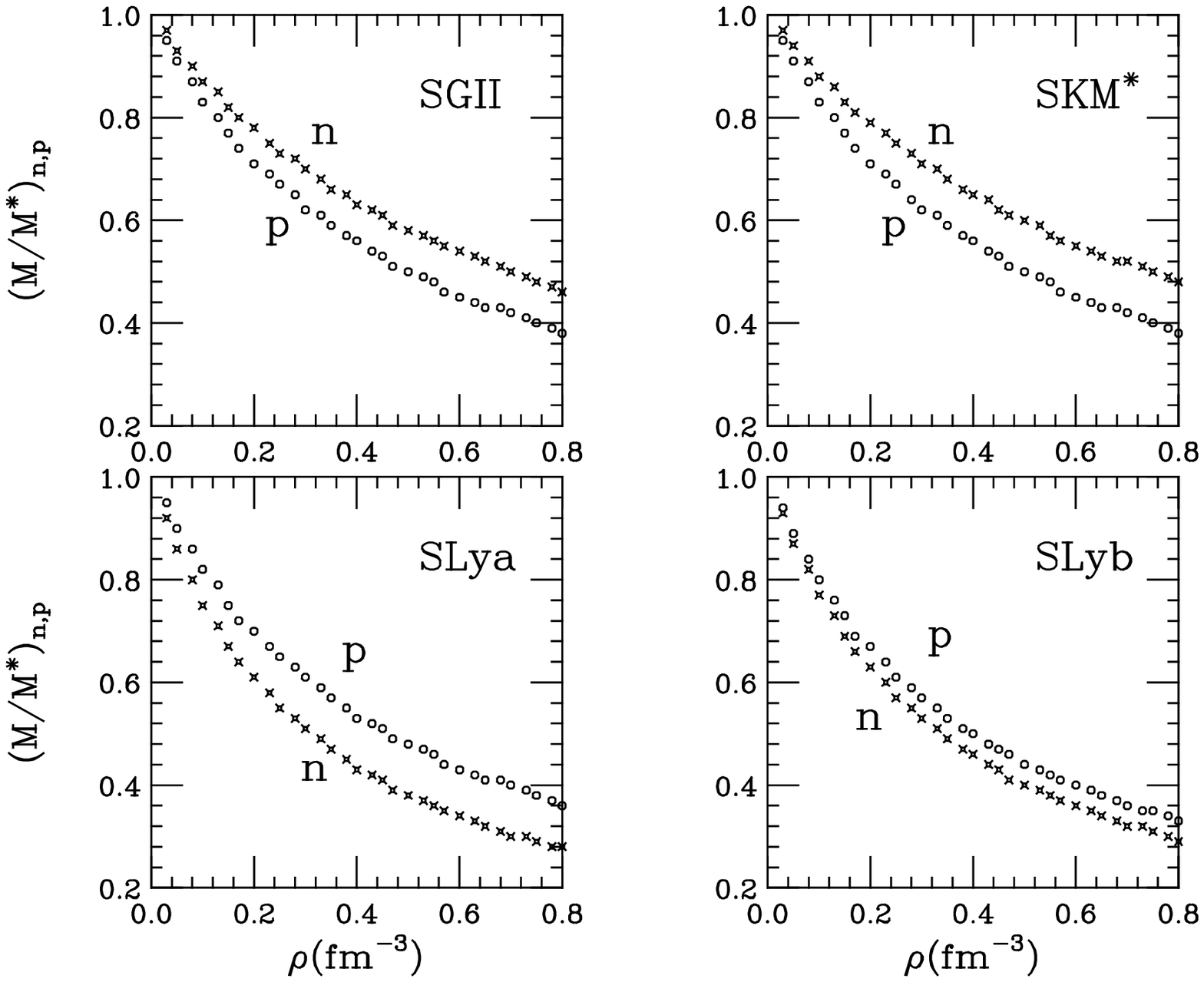}
\caption{\it Left: Density dependence of the potential symmetry term
for various Skyrme effective forces, see text.
The bottom right panel shows the kinetic contribution.
Right: Density dependence of the neutron/proton effective
mass splitting
for various Skyrme effective forces, see text.
The asymmetry is fixed at $I=0.2$, not very exotic.}
\label{fig:sksym}
\end{minipage}
\end{figure}

In the Fig. \ref{fig:sksym}(left) we show the density dependence
of the potential symmetry term of various Skyrme interactions, $SIII$,
 $SGII$, $SKM^*$ (see \cite{KrivineNPA336} and refs. therein) and the 
more recent Skyrme-Lyon forms, $SLya$ and $Slyb$ (or $Sly4$), 
see \cite{LyonNPA627,LyonNPA635}.
We also separately present the local and non-local contributions, first
and second term of the Eq.(\ref{sky}). We clearly see a sharp change
from the earlier Skyrme forces to the Lyon parametrizations, with 
almost an inversion of the signs of the two contributions, 
 \cite{BaranPR410}. The important
repulsive non-local part of the Lyon forces leads to a completely
different behavior of the neutron matter $EOS$, of great relevance 
for the neutron star properties. Actually this substantially modified
parametrization was mainly motivated by a very unpleasant feature
in the spin channel of the earlier Skyrme forces, the collapse of
polarized neutron matter, see discussion in 
\cite{KutscheraPLB325,LyonNPA627,LyonNPA635,LyonNPA665}.
In correspondence the
predictions on the isospin effects on the momentum
dependence of the symmetry term are quite different, see Fig.\ref{fig:sksym}
(left).
A very important consequence for the reaction dynamics is the
expected inversion of the sign of the $n/p$ effective mass splitting.

\subsection{Effective masses in neutron-rich matter}

In asymmetric matter we consistently have a splitting
of the neutron/proton effective masses given by:
\begin{equation}
{m^*_q}^{-1} = m^{-1} + g_1 \rho + g_2 \rho_q,
\label{skymass}
\end{equation}
with
$$
\rho_{q=n,p} = {{1 \pm I} \over 2} \rho~~~ (+,n).
$$
The $g_1,~g_2$ coefficients
are simply related to the momentum dependent part of the Skyrme forces:
\begin{eqnarray}
&& g_1 = {\frac {1}{4 \hbar^2}} [t_1 (2 + x_1) + t_2 (2 + x_2)] \nonumber\\ 
&& \nonumber\\
&& g_2 = {\frac {1}{4 \hbar^2}} [t_2 (1 + 2 x_2) - t_1 (1 + 2 x_1)] 
\label{g1g2}
\end{eqnarray}
This result derives from a general $q-structure$ of the momentum
dependent part of the Skyrme mean field
\begin{equation}
U_{q,MD} = m ( g_1 \rho + g_2 \rho_q) E
\label{momdep}
\end{equation}
where $E$ is the nucleon kinetic energy, while the total field seen by 
the $q$-nucleon has the form \cite{ColonnaPLB428}
\begin{equation}
U_q(\rho, \rho_q,k) = U_{q,MD} + {\frac {g_1 \hbar^2 \tau}{2}} +
 {\frac {g_2 \hbar^2 \tau_q}{2}} + {\frac  {\partial \epsilon_{Loc}(Pot)}
{ \partial \rho_q}}
\label{qmean}
\end{equation}

where $\hbar^2 \tau$ is the kinetic energy density and $\epsilon_{Loc}(Pot)$
the local part of the potential energy density.

In the Fig. \ref{fig:sksym}(right) we
show the density behavior of $m^*_{n,p}$ in  neutron rich
matter $I=0.2$ for the same effective interactions.
From the Eqs.(\ref{skymass}, \ref{g1g2}) we see that the sign of the
$g_2$ univocally assigns the sign of the splitting, i.e. $g_2<0$
gives larger neutron masses $m^*_n>m^*_p$ while we have the opposite 
for $g_2>0$.

In the Table \ref{skyg1g2} we report some results obtained with
various Skyrme forces for quantities of interest, around saturation, for the
present discussion. We show also the $E-slope$ of the corresponding
Lane Potential, see later, simply related to the isospin dependent part
of Eq.(\ref{momdep}).
For the effective mass parameters of Eq.(\ref{skymass})
we observe that while the $g_1$ coefficients are always positive, 
corresponding to a decrease of the
nucleon mass in the medium, the isospin dependent part shows different
signs.
In particular we see that in the Lyon forces the $g_2$ values are 
 positive, with neutron effective
masses below the proton ones for n-rich matter as shown in Fig.
\ref{fig:sksym} (right).

\vskip 0.5cm
\begin{table}
\caption{Properties at saturation}
\begin{tabular}{ c c c c c c c} \hline
$ Force$  & $ SIII$ &~ $ SGII$ &~$ SkM^*$ &~$SLya$ &~$SLy4$ &~$SLy7$ 
\\ \hline
$g_1~(10^{-3})(MeV^{-1}fm^3) $ & $+3.85$ & $+3.31$ & $+3.53$ & $+10^{-5}$ &
 $+1.67$ & $+1.70$ \\ \hline
$g_2~(10^{-3})(MeV^{-1}fm^3) $ & $-3.14$ & $-2.96$ & $-3.50$ & $+5.78$ &
 $+2.53$ & $+2.76$ \\ \hline
$\rho_0(fm^{-3})$ & $0.150$ & $0.1595$ & $0.1603$ & $0.160$ &
 $0.1595$ & $0.1581$ \\ \hline
$a_4(MeV)$ & $ 28.16$ & $ 26.83$ & $30.03$ & $31.97$ & $32.01$ & $32.01$ 
 \\ \hline
$C(\rho_0)(MeV)$ & $31.72$ & $29.06$ & $35.46$ & $39.40$ & $39.42$ & $39.42$ 
 \\ \hline
$E-slope(Lane Pot.)$ & $-0.22$ & $-0.21$ & $-0.26$ & $+0.43$ &
 $+0.19$ & $+0.20$ \\ \hline
\end{tabular}
\label{skyg1g2}
\end{table}

We note that the same is predicted
from microscopic relativistic Dirac-Brueckner calculations
\cite{HofmannPRC64,SchillerEPJA11,DalenNPA744} and in general from the 
introduction of 
scalar isovector
virtual mesons in $RMF$ approaches \cite{LiuboPRC65,GrecoPRC67}. At variance, 
non-relativistic Brueckner-Hartree-Fock
calculations are leading to opposite conclusions
 \cite{Bombiso,ZuoPRC60}. We remind that a comparison between relativistic
effective ($Dirac$) masses and non-relativistic effective masses requires
some attention. This point will be carefully discussed later.

\subsection{Energy dependence of the Lane Potential}
The sign of the splitting will directly affect the energy
dependence of the Lane Potential, i.e. the difference between $(n,p)$ optical
potentials on charge asymmetric targets, normalized by the target asymmetry
\cite{LaneNP35}.  
From the Eqs.(\ref{momdep},\ref{qmean}) we obtain the explicit Skyrme
form of the Lane Potential:
\begin{equation}
U_{Lane} \equiv {\frac {U_n - U_p}{2I}} = C(\rho_0) - {2 \over 3}
m \rho_0 \epsilon_F \Big[ g_1 + {7 \over 12}g_2 \Big] + 
 {\frac{m\rho_0}{2}} g_2 E
\label{lane}
\end{equation}
where $C(\rho_0)$ gives the potential
part of the $a_4$ parameter in the mass formula.
We see that the $E-slope$ has just the sign of the $g_2$ parameter, and so
we have opposite predictions from the various Skyrme forces analysed here. 
The change
in the energy slope is reported in the last row of the Table \ref{skyg1g2}.
The difference in the energy dependence of the Lane Potential is quite
dramatic.

The second term of the Eq.(\ref{lane}) is also interesting. In the case
$g_2>0$ (positive slope) decreases the starting zero energy point,
while in the case $g_2<0$ (negative slope) it represents a small correction
to the symmetry energy $C(\rho_0)$. We then expect to see a $crossing$
of the two prescriptions at very low energies, i.e. low momentum
nucleons will see exactly the same Lane potentials.

An important physical consequence of the negative slopes is that the isospin
effects on the optical potentials tend to disappear at energies just above
$100~MeV$ (or even change the sign for ``old'' Skyrme-like forces).
Unfortunately results derived from neutron/proton optical potentials at 
low energies
are not conclusive, \cite{LaneNP35,BecchettiPR182,Hodopt}, since the 
effects appear
of the same order of the uncertainty on the determination of the local
contribution. Moreover at low energies we expect the crossing discussed
before therefore for a wide energy range we cannot see differences.
More neutron data are needed at higher energies, in particular 
a systematics of the energy dependence. 

We can expect important effects
on transport properties ( fast particle emission, collective flows)
 of the dense and asymmetric $NM$ that will
be reached in Radioactive Beam collisions at intermediate energies,
i.e. in the $RIA$ energy range.

\section{Symmetry energy in Quantum-Hadro-Dynamics}

The $QHD$ effective field model represents a very successful attempt to 
describe, in a fully relativistic picture, equilibrium and 
dynamical properties of nuclear systems at the hadronic 
level \cite{WaleckaAP83,SerotANP16,SerotIJMPE6}.
Here we focus on the dynamical response and static (equilibrium) 
properties of 
Asymmetric Nuclear Matter ($ANM$) in a ``minimal'' effective meson
field model, with two meson (scalar and vector) contributions
in each isospin channel, $(\sigma,\omega)$ for the isoscalar
and $(\delta,\rho)$ for the isovector part, see
details in \cite{BaranPR410} and refs. therein. In particular
we will discuss the effects of the isovector/scalar $\delta$ meson,
 since this is directly related to the nucleon effective mass
splitting of interest here.

The symmetry energy in $ANM$ is defined from the expansion
of the energy per nucleon $E(\rho_B,I)$ in terms of the asymmetry
parameter 
$I \equiv -\frac{\rho_{B3}}{\rho_B}=\frac{\rho_{Bn}-\rho_{Bp}}
{\rho_B}=\frac{N-Z}{A}$.
We have
\begin{equation}\label{eq.7}
E(\rho_B,I)~\equiv~\frac{\epsilon(\rho_B,I)}{\rho_B}~
=~E(\rho_B) + E_{sym}(\rho_B) I^2 \nonumber \\
+ O(I^4) +...
\end{equation}
and so in general
\begin{equation}\label{eq.8}
E_{sym}~\equiv~\frac{1}{2} \frac{\partial^2E(\rho_B,I)}
{\partial I^2} \vert_{I=0}~=~\frac{1}{2} \rho_B 
\frac{\partial^2\epsilon}{\partial\rho_{B3}^2}
 \vert_{\rho_{B3}=0}
\end{equation}

In the Hartree case an explicit expression for the symmetry energy is 
easily derived 
that can be reduced to a very simple form,  with transparent  
$\delta$-meson effects \cite{LiuboPRC65,GrecoPRC67}:
\begin{equation}\label{eq.11}
E_{sym}(\rho_B) = \frac{1}{6} \frac{k_F^2}{E^*_F} + \frac{1}{2}
\left[f_\rho - f_\delta 
\left(\frac{m^*}{E^*_F}\right)^2\right] \rho_B
\end{equation}

where $k_F$ is the nucleon Fermi momentum corresponding to $\rho_B$, 
$E^*_F \equiv \sqrt{(k_F^2+{m^*}^2)}$ and $m^*$ is the effective
nucleon mass in symmetric $NM$, $m^*=m + \Sigma_s(\sigma)$,
with the isoscalar, scalar self-energy 
$\Sigma_s(\sigma) \equiv -f_{\rho} \rho_s$ ($\rho_s$ scalar density).
$f_i \equiv ({g_i \over m_i})^2,~i=\sigma, \omega, \delta, \rho),$
 are the effective meson coupling constants \cite{par}.

We see that, when the $\delta$ is included, the empirical $a_4$ value actually
corresponds to the combination $[f_\rho - f_\delta (\frac{M}{E_F})^2]$
of the $(\rho,\delta)$ coupling constants. Therefore if
$f_\delta \not= 0$ we have to increase correspondingly the $\rho$-coupling. 

\begin{figure}[htb]
\begin{minipage}[t]{67mm}
\includegraphics*[scale=0.45]{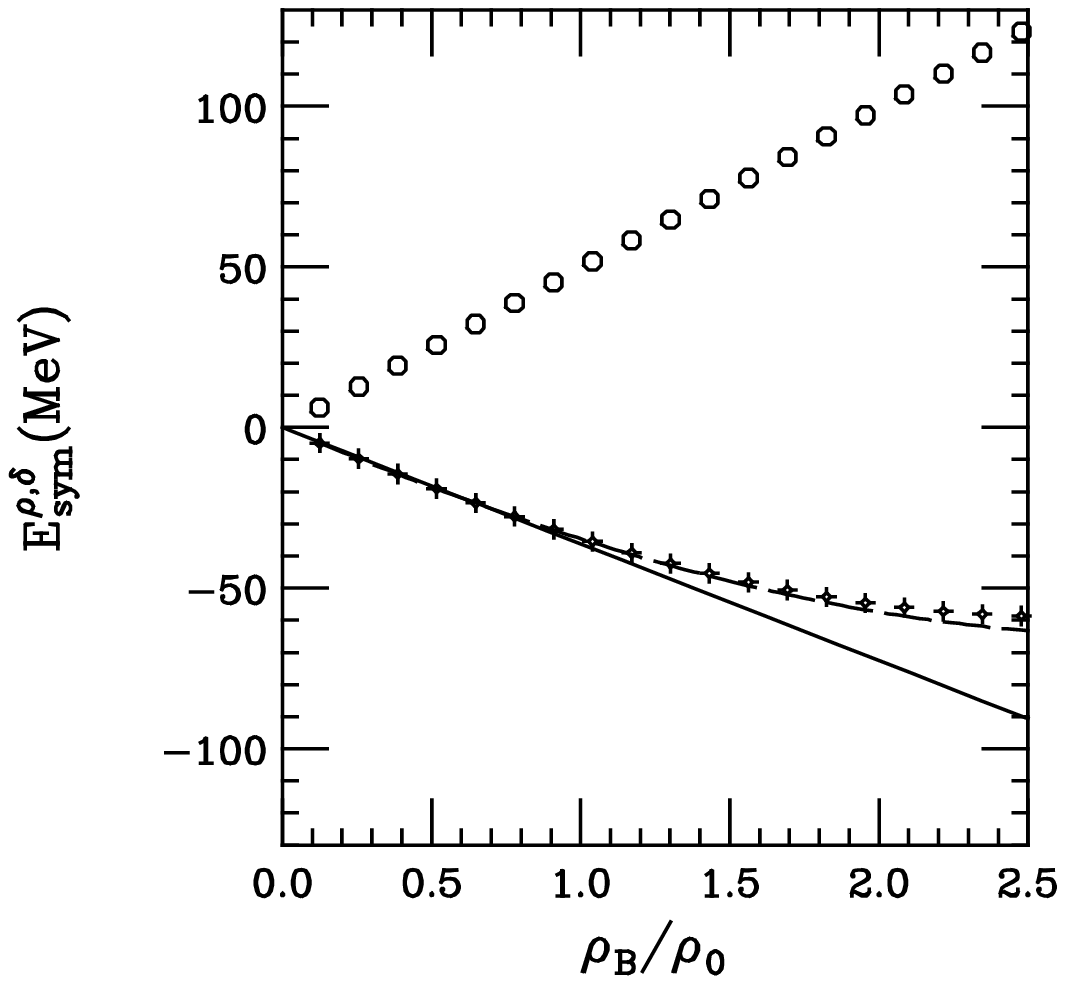}

\end{minipage}
\hspace{\fill}
\begin{minipage}[t]{67mm}
\includegraphics*[scale=0.40]{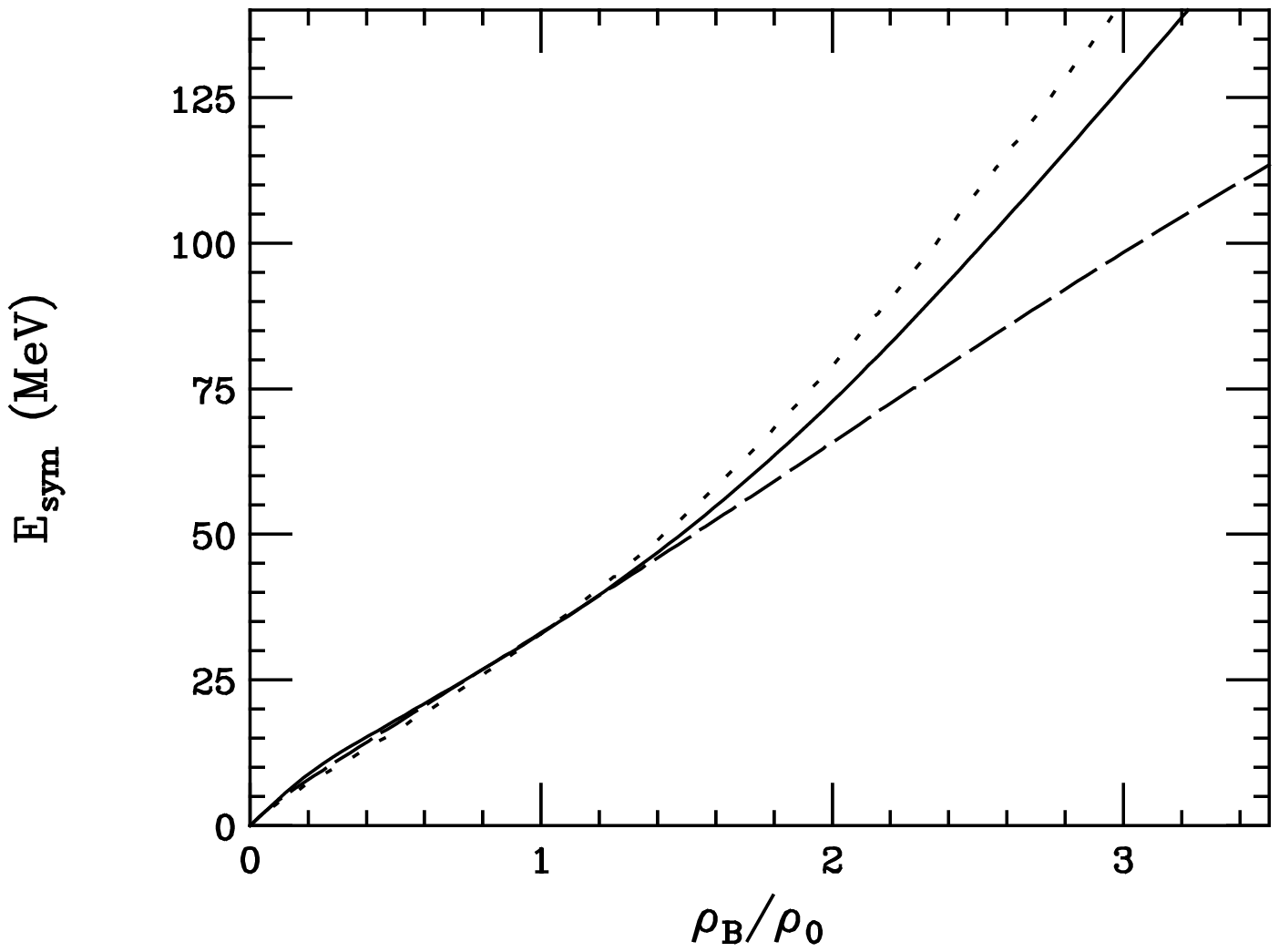}
\caption{\it Left: $\rho$- (open circles) and $\delta$- (crosses) contributions
to the potential symmetry energy.
The solid line is a linear extrapolation of the 
low density $\delta$ contribution.
Right: Total (kinetic + potential) symmetry energy as a function of the
baryon density. Dashed line ($NL\rho$). Dotted line ($NL\rho\delta$).
Solid line $NLHF$, only Fock correlations \cite{GrecoPRC67}.}
\label{fig:qhdsym}
\end{minipage}
\end{figure}

Now the symmetry energy at 
saturation
density is actually built from the balance of scalar (attractive) and vector
(repulsive) contributions, with the scalar channel becoming weaker 
with increasing
baryon density. This is clearly shown in
Fig.\ref{fig:qhdsym}(left). This is indeed the isovector 
counterpart of the saturation
mechanism occurring in the isoscalar channel for symmetric nuclear matter.

In  Fig.\ref{fig:qhdsym}(right) we show the total symmetry energy for the 
different models.
At subnuclear densities, $\rho_B < \rho_0$, in both cases, 
$NL\rho$ and $NL\rho\delta$, from Eq.(\ref{eq.11}) we have an almost 
linear dependence 
of $E_{sym}$ on
the baryon density, since $m^* \simeq E_F$ as a  good approximation .
Around and above $\rho_0$ we see a steeper increase in the
$(\rho+\delta)$ case since $m^*/E_F$ is decreasing. 

In conclusion when the $\delta$-channel is included the behaviour 
of the symmetry energy
is stiffer at high baryon density 
from the relativistic
mechanism discussed before. This is in fact due to a larger
contribution from the $\rho$ relative to the $\delta$ meson. We expect to
see these effects more clearly in the relativistic reaction dynamics at
intermediate energies, where higher densities are reached.

\subsubsection{Nucleon Effective Mass Splitting}

An important qualitatively new result of the $\delta$-meson coupling is
the $n/p$-effective mass splitting in asymmetric matter, 
\cite{BaranPR410} and refs. therein, $Dirac$ mass, see later:
\begin{equation}\label{diracms}
m^*_D(q)= m + \Sigma_s(\sigma) \pm f_\delta \rho_{S3}~, 
\end{equation} 
where $+$ is for neutrons. Since $\rho_{S3} \equiv \rho_{Sp} - \rho_{Sn}$,
 in $n$-rich 
systems we have a neutron $Dirac$ effective
mass always smaller than the proton one.  
We note again that a decreasing neutron 
effective mass in $n$-rich matter is a direct consequence of the 
relativistic mechanism
for the symmetry energy, i.e. the balance of scalar (attractive) 
and vector (repulsive)
contributions in the isovector channel.

\subsection{Dirac and Schr\"odinger Nucleon Effective Masses in 
Asymmetric Matter}
The prediction of a definite $m^*_D(n)<m^*_D(p)$ effective mass splitting
in $RMF$ approaches, when a scalar $\delta-like$ meson is included,
is an important result that requires some further analysis, in 
particular relative to the controversial predictions of non-relativistic
approaches.

Here we are actually discussing the {\it Dirac} effective masses 
$m^*_D(n,p)$,
i.e. the effective mass of a nucleon in the in-medium Dirac equation
with all the meson couplings. The relation to the {\it Schr\"odinger}
effective masses $m^*_S(n,p)$, i.e. the ``k-mass'' due to the momentum 
dependence
of the mean field in the non-relativistic in-medium Schr\"odinger
equation is not trivial, see 
\cite{BouyssyPRC36,SerotANP16,JaminonPRC22,JaminonPRC40}.
We will extend the argument of the refs.
\cite{JaminonPRC22,JaminonPRC40} to the case of
asymmetric matter.

We start from the simpler symmetric case without self-interacting terms.
The nucleon Dirac equation in the medium contains the scalar
self-energy $\Sigma_s = -f_\sigma \rho_S$ and the vector self-energy
(fourth component) $\Sigma_0 = f_\omega \rho_B$ and thus the
corresponding energy-momentum relation reads:
\begin{equation}\label{enmom}
(\epsilon + m - \Sigma_0)^2 = p^2 + (m + \Sigma_s)^2 = p^2 + {m_D^*}^2 
\end{equation}
i.e. a dispersion relation
\begin{equation}\label{disp}
\epsilon = - m + \Sigma_0 + \sqrt{p^2 + {m_D^*}^2} 
\end{equation}

From the total single particle energy $E = \epsilon + m$ expressed in the form
$E = \sqrt{k_{\infty}^2 + m^2}$, where $k_{\infty}$ is the relativistic
asymptotic momentum, using  Eq.(\ref{enmom}) we can get the relation
\begin{eqnarray}\label{ueff}
\frac{k_{\infty}^2}{2m} = \epsilon + \frac{\epsilon^2}{2m} =
 ~~~~~~~~~~~~~~~ ~~~~~~~~~~~~~~~ ~~~~~~~~~~~~~~~ \nonumber\\
  \frac{p^2}{2m} + \Sigma_s + \Sigma_0 + \frac{1}{2m}(\Sigma_s^2 - \Sigma_0^2)
 + \frac{\Sigma_0}{m}\epsilon \equiv \frac{p^2}{2m} + 
 U_{eff}(\rho_B,\rho_s,\epsilon)
\end{eqnarray}
i.e. a Schr\"odinger-type equation with a momentum dependent mean field
that with the dispersion relation Eq.(\ref{disp}) is written as
\begin{eqnarray}\label{unonrel}
U_{eff} =  \Sigma_s + \frac{1}{2m}(\Sigma_s^2 + \Sigma_0^2)
 + \frac{\Sigma_0}{m}\sqrt(p^2 + {m_D^*}^2) \nonumber\\
 \simeq \Sigma_s + \frac{\Sigma_0 m_D^*}{m} + 
 \frac{1}{2m}(\Sigma_s^2 + \Sigma_0^2) + \frac{p^2}{2m}\frac{\Sigma_0}{m_D^*}
\end{eqnarray}
The relation between Schr\"odinger
and Dirac nucleon effective masses is then
\begin{equation}\label{effmas}
m_{S}^* = \frac{m}{1 + \frac{\Sigma_0}{m_D^*}} = m_D^* \frac{m}{m + 
 \Sigma_s + \Sigma_0}
\end{equation}
Since at saturation the two self-energies are roughly compensating each other,
$\Sigma_s + \Sigma_0 \simeq -50MeV$ the two effective masses
are not much different, with the $S-mass$ slightly larger than the $D-mass$.




In the case of asymmetric matter, neutron-rich as always considered here,
we can have two cases:
\begin{itemize}
\item {\it Only $\rho~meson$ coupling}

Now the scalar part is not modified, we have the same scalar self energies
$\Sigma_s$ for neutrons and protons and so the same Dirac masses. The vector
self energies will show an isospin dependence with a new term
$\mp f_{\rho}\rho_{B3}$, repulsive for neutrons ($-$ sign, since we
use the definition $\rho_{(B,S)3} \equiv \rho_{(B,S)p}-\rho_{(B,S)n}$).
As a consequence we see a splitting at the level of the
Schr\"odinger masses since Eq.(\ref{effmas}) becomes 

\begin{equation}\label{effmasrho}
{m_{S}^*}(n,p) = m_D^* \frac{m}{m + 
 (\Sigma_s + \Sigma_0)_{sym} \mp f_{\rho}\rho_{B3}} 
\end{equation}

in the direction of $m_S(n)^*<m_S(p)^*$ (here and in the following 
upper signs are for neutrons). 

\item{\it $\rho+\delta$ coupling}

The above splitting is further enhanced by the direct effect of the
scalar isovector coupling, see Eq.(\ref{diracms}). Then the
Schr\"odinger masses are

\begin{equation}\label{effmasrhodelta}
{m_{S}^*}(n,p) = 
({m_D^*}_{sym} \pm f_{\delta}\rho_{S3}) \frac{m}{m + 
 (\Sigma_s + \Sigma_0)_{sym} \mp (f_{\rho}\rho_{B3} - f_{\delta}\rho_{S3})}
\end{equation}

The new term in the denominator will further contribute to the 
$m^*_S(n)<m^*_S(p)$
splitting since we must have $f_{\rho}>f_{\delta}$ in order to get
a correct symmetry parameter $a_4$. The effect is larger at higher
baryon densities because of the decrease of the scalar $\rho_{S3}$, due
to the faster $\frac{m_{Dn}^*}{E_{Fn}^*}$ reduction of $\rho_{Sn}$.
\end{itemize}

Actually in a non-relativistic limit we can approximate in Eq.(\ref{ueff})
directly the energy $\epsilon$ with the asymptotic kinetic energy
leading to the much simpler relation:

\begin{equation}\label{nreleffmas}
m_{S}^* = \frac{m}{1 + \frac{\Sigma_0}{m}} \simeq {m - \Sigma_0}  
 = m_D^* - (\Sigma_s + \Sigma_0).
\end{equation}

In the case of asymmetric matter this leads to the more transparent
relations
\begin{itemize}
\item {\it Only $\rho~meson$ coupling}

\begin{equation}\label{nrelmasrho}
{m_{S}^*}(n,p) = m_D^* - (\Sigma_s + \Sigma_0)_{sym} \pm f_{\rho}\rho_{B3}
\end{equation}

\item{\it $\rho+\delta$ coupling}

\begin{equation}\label{nrelmasrhodelta}
{m_{S}^*}(n,p) = 
{m_D^*}_{sym} - (\Sigma_s + \Sigma_0)_{sym}
\pm (f_{\rho}\rho_{B3} - f_{\delta}\rho_{S3})
\end{equation}

\end{itemize}

In conclusion any $RMF$ model will predict
the definite isospin splitting of the nucleon effective masses 
$m_S(n)^*<m_S(p)^*$
in the non-relativistic limit. We expect an increase of the difference
of the neutron/proton mean field at high momenta, with important
dynamical contributions that will enhance the transport effects of the 
symmetry energy.

\subsubsection{Beyond RMF: k-dependence of the Self-Energies}

Correlations beyond the Mean Field picture will lead to a further
density and, most important, momentum dependence of the effective
couplings and consequently of the Self-Energies. This can be clearly seen
from microscopic Dirac-Brueckner ($DBHF$) calculations,
see \cite{HofmannPRC64,SchillerEPJA11,DalenNPA744} and refs. therein.
Even the basic Fock correlations are implying a density dependence
of the couplings, see \cite{GrecoPRC67}. In fact from the very first
relativistic transport calculations a momentum dependence of the vector fields
was required in order to reduce the too large repulsion of the otpical 
potential at high momenta, see \cite{GiessenRPP56}.
From the above discussion we can expect that, when the $\delta$-meson is 
included \cite{delta-corr}, the sign of the $Dirac$-mass splitting will 
not be modified,i.e. $m_{Dn}^*<m_{Dp}^*$. In fact this is the result of 
the microscopic
$DBHF$ calculations in asymmetric matter of 
refs.\cite{HofmannPRC64,SchillerEPJA11,DalenNPA744}.
The problem is however that in correspondence the Schr\"odinger
mass splitting can have any sign. Indeed this is the case of the very
recent $DBHF$ analysis of the T\"ubingen group \cite{DalenNPA744,Dalenar0502}.
For the $S$-masses they get an opposite behavior  $m_{Sn}^*>m_{Sp}^*$, with
large fluctuations around the Fermi momenta, due to the opening of the phase 
space for intermediate inelastic channels. These results are actually
dependent on the way the self-energies (mean fields) are derived from the
full correlated theory and different predictions can be obtained,
see the recent refs.\cite{MaPLB604,Sammar0301,Sammar0411}.

All that clearly shows that experimental observables are in any case
strongly needed.

\subsection{The Dirac-Lane Potential}
In the non-relativistic limit of the Eq.(\ref{ueff}) we can easily extract 
the neutron-proton mean optical potential using the isospin dependence
of the self-energies

\begin{eqnarray}\label{npself}
\Sigma_{0,q} = \Sigma_{0,sym} \mp f_{\rho}\rho_{B3}~,
\nonumber \\
\Sigma_{s,q} = \Sigma_{s,sym} \pm f_{\delta}\rho_{S3}~, 
\end{eqnarray}

With some algebra we get a compact form of the $Dirac-Lane~Potential$

\begin{eqnarray}\label{diraclane}
{U_{Dirac-Lane}} \equiv \frac{U_n-U_p}{2I} = \nonumber \\
\rho_0 \Big[ f_{\rho} (1- \frac{\Sigma_{0,sym}}{m}) - 
f_{\delta}{\frac{\rho_{S3}}{\rho_{B3}}}(1+ \frac{\Sigma_{s,sym}}{m})\Big]
 + f_{\rho} \frac{\rho_0}{m} \epsilon
\end{eqnarray}  

It is interesting to compare with the related discussion presented
 for the non-relativistic effective forces, mainly of
Skyrme-like form.
First of all we predict a definite positive $E-slope$ given by 
the quantity $f_{\rho} \frac{\rho_0}{m}$, which is actually not large
within the simple $RMF$ picture described here. This is a obvious
consequence of the fact that the ``relativistic'' mass splitting is always
in the direction $m_S(n)^*<m_S(p)^*$.
The realistic magnitude of the effect could be different
 if we take into account that
some explicit momentum dependence should be included in the scalar and
vector self energies, \cite{GiessenRPP56}, as also discussed before.
An interesting point in this direction comes from the phenomenological
Dirac Optical Potential ($Madland-potential$) constructed in the refs.
\cite{KozackPRC39,KozackNPA509}, fitting simultaneously proton and
neutron (mostly total cross sections) data for collisions with a wide
range of nuclei at energies up to $100~MeV$. Recently this Dirac optical
potential has been proven to reproduce very well the new neutron 
scattering data on $^{208}Pb$ at $96~MeV$ \cite{KlugPRC6768} measured at
the Svendberg Laboratory in Uppsala.

\begin{figure}
\includegraphics*[scale=0.35]{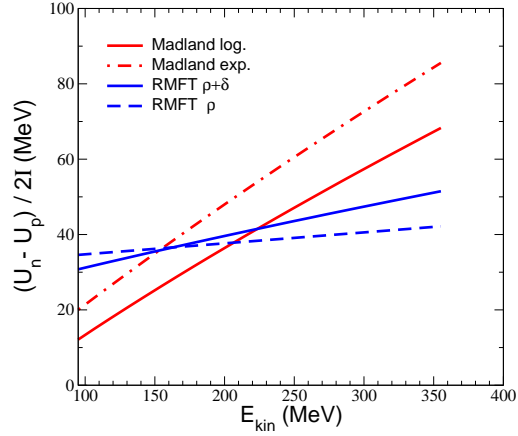}
\caption{\it
Energy dependence of the Dirac-Lane potential in the $RMF$ picture
(solid: $NL\rho$; dashed: $NL\rho\delta$) and in the phenomenologic
Dirac Optical Model of Madland et al. \cite{KozackPRC39,KozackNPA509},
see text.}
\label{dlane}
\end{figure}

The phenomenological $Madland-potential$ has different implicit 
momentum dependences 
($exp/log$) in the self-energies, \cite{KozackPRC39,KozackNPA509}.
In Fig.\ref{dlane} we show the corresponding energy dependences for the
$Dirac-Lane$ potentials, compared to our $NL\rho$ and $NL\rho\delta$
estimations. When we add the $\delta$-field we have a larger slope since
the $\rho-coupling$ should be increased.
The slopes of the phenomenological potentials are systematically
larger, interestingly similar to the ones of the $Skyrme-Lyon$ forces. 
We note that in the $Madland-potential$
the Coulomb interaction is included, i.e. an extra repulsive vector 
contribution for the protons.
The first, not energy dependent, term of Eq.(\ref{diraclane}) is directly
related to the symmetry energy at saturation, exactly like in the
non-relativistic case, see Eq.(\ref{lane}).

The conclusion is that a good {\it systematic} measurement of the Lane 
potential in a wide
range of energies, and particularly around/above $100~MeV$,
would answer many fundamental questions in isospin physics.

\section{Reaction Observables}

In order to directly test the influence of the $S$-effective mass
splitting we will present results from reaction dynamics at intermediate
energies analysed in a non-relativistic transport approach, of $BNV$
type, see \cite{BaranPR410}. The $Iso-MD$ effective interaction is
derived via an asymmetric extension of the $GBD$ force, 
 \cite{GalePRC41,GrecoPRC59},
constructed in order to have a strict correspondence to the Skyrme 
forces, see also \cite{RizzoNPA732}.

\begin{figure}[htb]
\centering
\begin{picture}(2,2)
\put(105.5,40.6){\mbox{\includegraphics[width=3.5cm]{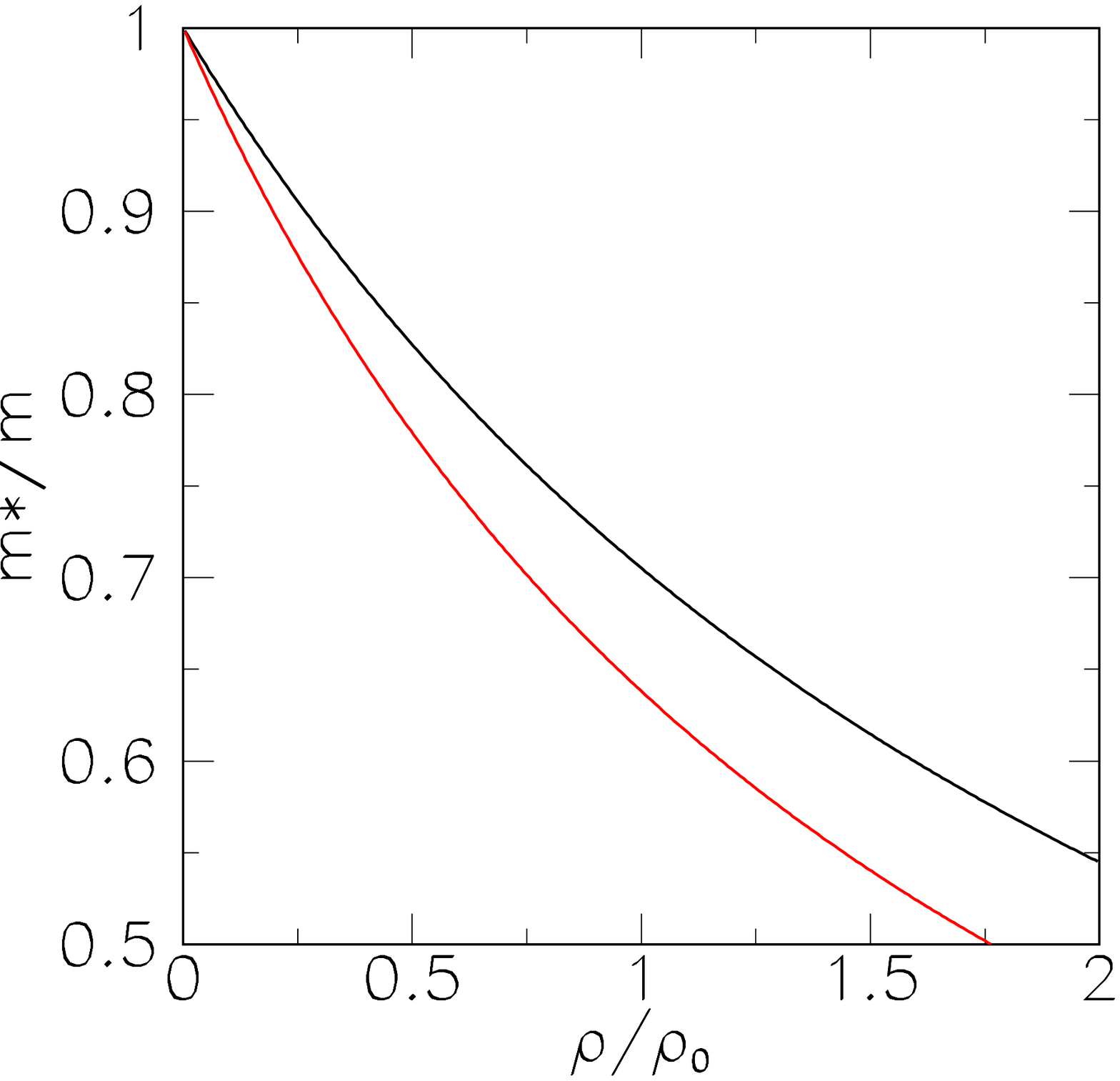}}}
\end{picture}
\includegraphics[width=7cm]{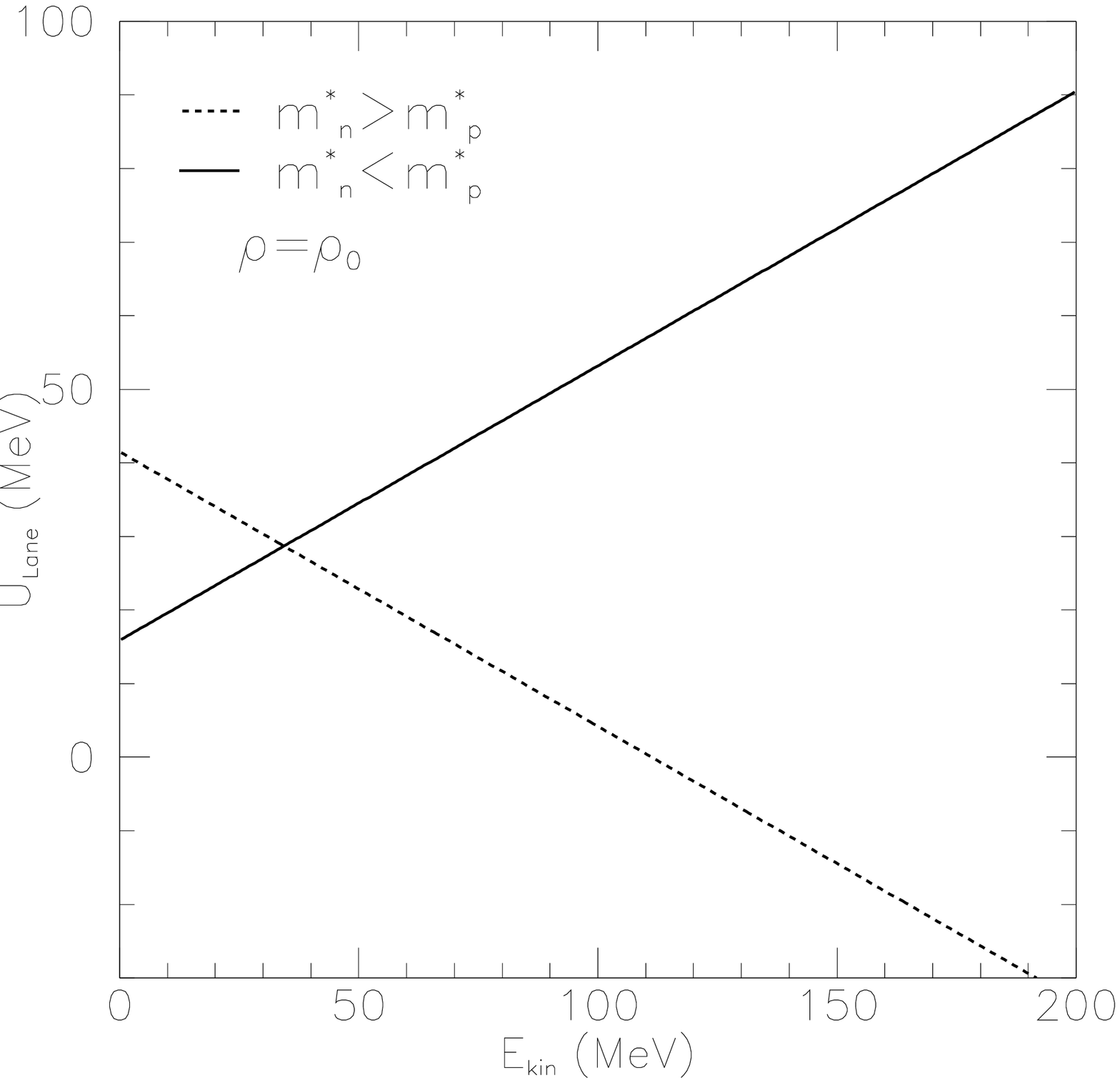}

\caption{\it Lane potential for the used parametrizations; 
small panel: the effective mass splitting, related to the slope of the Lane
potential.} 
\label{lanepseudo}
\end{figure}

The energy density  can be parametrized
as follows:
\begin{equation}
\varepsilon=\varepsilon+\varepsilon_{kin}+\varepsilon(A',A'')
+\varepsilon(B',B'')+\varepsilon(C',C'')
\label{edensmd1}
\end{equation}
where $\varepsilon_{kin}$ is the usual kinetic energy density and
\begin{eqnarray}
\varepsilon(A',A'')=(A'+A''\beta^2)\frac{\varrho^2}{\varrho_0}
 \nonumber \\
\varepsilon(B',B'')=(B'+B''\beta^2)\left ( \frac{\varrho}{\varrho_0}
\right )^{\sigma}\varrho
 \nonumber \\
\varepsilon(C',C'')=C'(\mathcal{I}_{NN}+\mathcal{I}_{PP})
+C''\mathcal{I}_{NP}
\label{edensmd2}
\end{eqnarray}
Here $\mathcal{I}_{\tau \tau'}$ are integrals of the form:
$$\mathcal{I}_{\tau \tau'}=\int d \vec{p} \; d \vec{p}\,' 
f_{\tau}(\vec{r},\vec{p})
 f_{\tau'}(\vec{r},\vec{p}\,') g(\vec{p},\vec{p}\,')$$

with $g(\vec{p},\vec{p}\,')=g(\vec{p}-\vec{p}\,')^2$
This choice of the function $g(\vec{p},\vec{p}\,')$ corresponds to
a Skyrme-like behaviour and it is suitable for $BNV$ simulations. 
In this frame
we can easily adjust the parameters in order to have the same density
dependence of the symmetry energy 
{\it but with two opposite n/p effective mass splittings}. So we can 
separately study the correspondent dynamical effects, \cite{RizzoNPA732}.
In Fig. \ref{lanepseudo} we show the $E$-dependence of the Lane potentials
with the parametrizations  corresponding to the two choices of the sign
of the n/p mass splitting (shown in the insert for the $I=0.2$ asymmetry).
The value of the splitting is exactly the same, only the sign is opposite.

The upper curve well reproduces the Skyrme-Lyon (in particular $SLy4,Sly7$)
results. The lower (dashed) the $SIII,SKM^*$ ones. We clearly see the crossing
at low energy, as expected from the previous discussion.

From the figure we can immediately derive the expectation of very different
symmetry effects for nucleons around $100MeV$ kinetic energy,
enhancement (with larger neutron repulsion) in the  $m_n^*<m_p^*$ case
vs. a  disappearing (and even larger proton repulsion) in the
$m_n^*>m_p^*$ choice. So it seems naturai to look at observables
where neutron/proton mean fields at high momentum are playing an
important role. We will show results for fast nucleon emissions and
collective flows at intermediate energies, in particular for
high transverse momentum selections.

\subsection{Isotopic content of the fast nucleon emission}

\begin{figure}[htb]
\centering
\includegraphics[width=8cm]{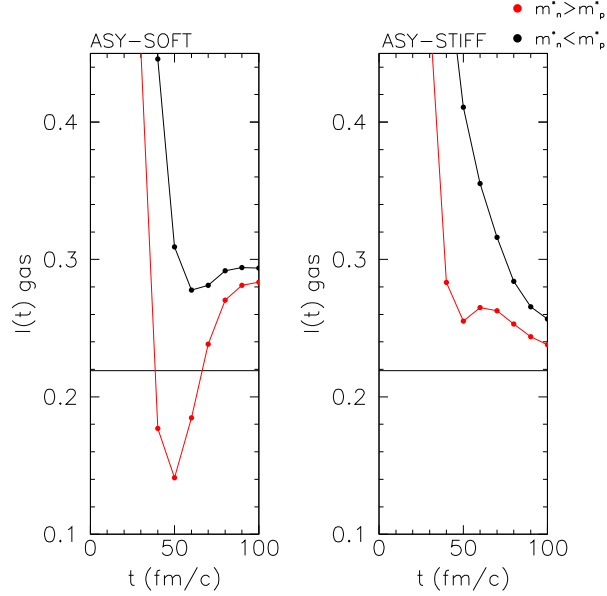}
\caption{\it Isospin gas content as a function of time in a central collision 
$^{132}Sn+^{124}Sn$ at $50 \,AMeV$ for two opposite choices of mass splitting.
Left: asy-soft EOS; right: asy-stiff EOS.}
\label{asyevol}
\end{figure}

We have performed realistic ``ab initio'' simulations of collisions
at intermediate energies of n-rich systems, in particular 
$^{132}Sn+^{124}Sn$ at $50~and~100, AMeV$, central selection.
Performing a local low density selection of the test particles
($\rho<\rho_0/8$) we can follow the time evolution of nucleon
emissions (gas phase) and the corresponding asymmetry.
In Fig. \ref{asyevol} we show the evolution of the
gas asymmetry ($I(t) \equiv (N-Z)/A$) for the $50AMeV$ reaction
 (the solid line gives the initial asymmetry).
For both choices of the density stiffness of the symmetry
term we clearly see the effects of the mass splitting, resulting in
a reduced fast neutron emission when $m_n^*>m_p^*$. 

\begin{figure}[htb]
\centering
\includegraphics[width=8cm]{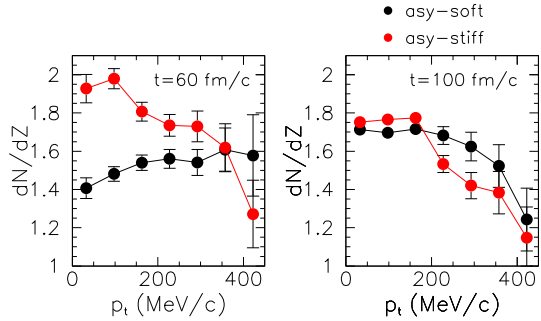}
\caption{\it Transverse momentum dependence of the neutron/proton
ratio in the rapidity range $|y^{(0)}| \leq 0.3$ for a central reaction 
$^{132}Sn+^{124}Sn$ at $50 \,AMeV$. Comparison between two choices of symmetry
energy stiffness at different times.}  
\label{sn132_50}
\end{figure}

\begin{figure}[htb]
\centering
\includegraphics[width=9cm]{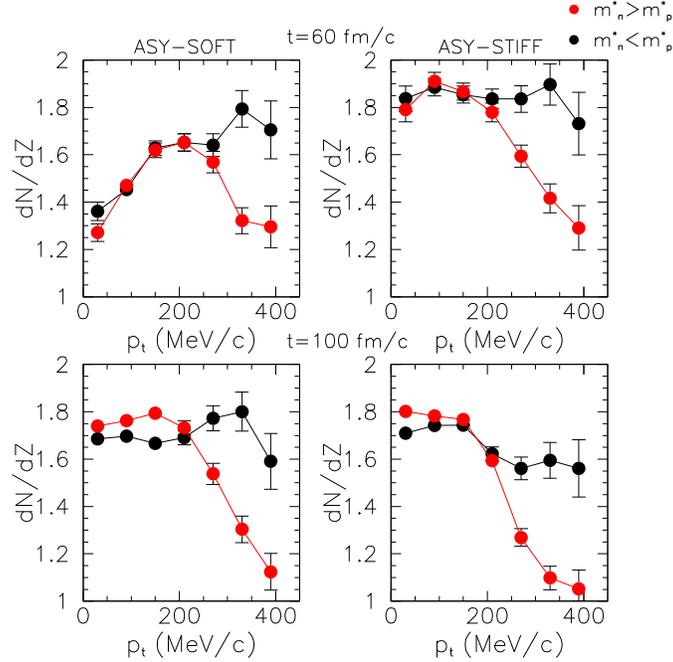}
\caption{\it Same as Fig. \ref{sn132_50}, for two opposite choices of mass
splitting.} \label{sn132_50_split}
\end{figure}

In order to better disentangle the mass splitting effect and to select
the corresponding observables, in Figs. \ref{sn132_50}, \ref{sn132_50_split}
 we report the $N/Z$ of the ``gas'' at two different times, $t=60fm/c$,
 end of the pre-equilibrium emission, and $t=100fm/c$ roughly freeze out 
time. We have followed the transverse momentum dependence, for
 a  fixed central rapidity $y^(0)$ (normalized to projectile rapidity).
In Fig. \ref{sn132_50} we show the results {\it without the mass splitting
effect}, i.e. only taking into account the different repulsion of the
symmetry term (the intial average asymmetry is $N/Z=1.56$).

At early times (left panel of Fig. \ref{sn132_50}), when the emission 
from high density regions is
dominant, we see a difference due to the larger neutron repulsion
of the the asy-stiff choice. The effect is reduced at higher $p_t$s
due to the overall repulsion of the isoscalar momentum dependence.
Finally at freeze out (right panel)
the difference is almost disappearing even for the more efficient
$Isospin~Distillation$ of the asy-soft choice during the expansion phase, 
 \cite{BaranPR410} and refs. therein.

When we introduce the mass splitting the difference in the isotopic content
of the gas, for the larger transverse momenta, is very evident at all times,
 in particular at the freeze-out of experimental interest, 
see Fig. \ref{sn132_50_split}.
As expected the $m_n^*>m_p^*$ sharply reduces the neutron emission
at high $p_t$s.

\begin{figure}[htb]
\centering
\includegraphics[width=9cm]{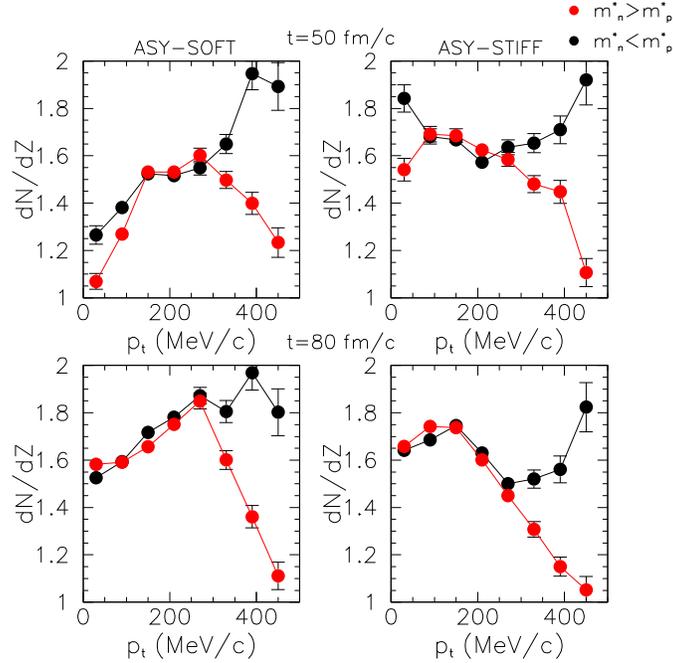}
\caption{\it Same as Fig. \ref{sn132_50_split}, but for an incident energy of 
$100\, AMeV$.} \label{sn132_100_split}
\end{figure}

We have repeated the analysis at higher energy, $100AMeV$, and the
effect appears nicely enhanced, see Fig. \ref{sn132_100_split}.

Similar results have been obtained in ref. \cite{BaoNPA735} at $400AMeV$
in a different $Iso-MD$ model, restricted to the $m_n^*>m_p^*$ choice.
In conclusion it appears that the isospin content of pre-equilibrium
emitted nucleons at high transverse momentum can solve the
mass-splitting puzzle.

\subsection{Differential collective flows and mass splittings}

\begin{figure}[htb]
\centering
\begin{picture}(0,0)
\put(150.0,4.8){\mbox{\includegraphics[width=4.2cm]{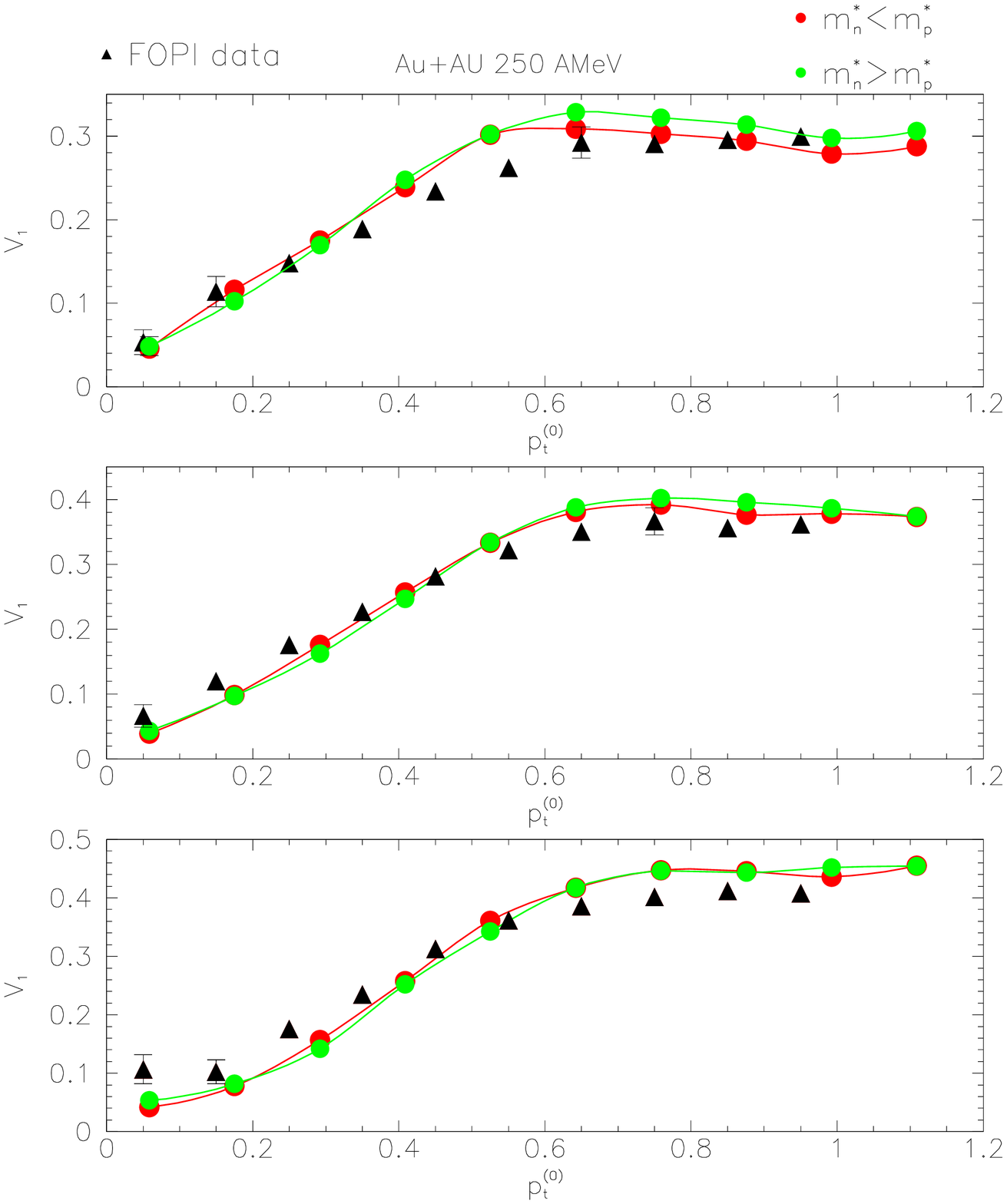}}}
\end{picture}
\includegraphics[width=9cm]{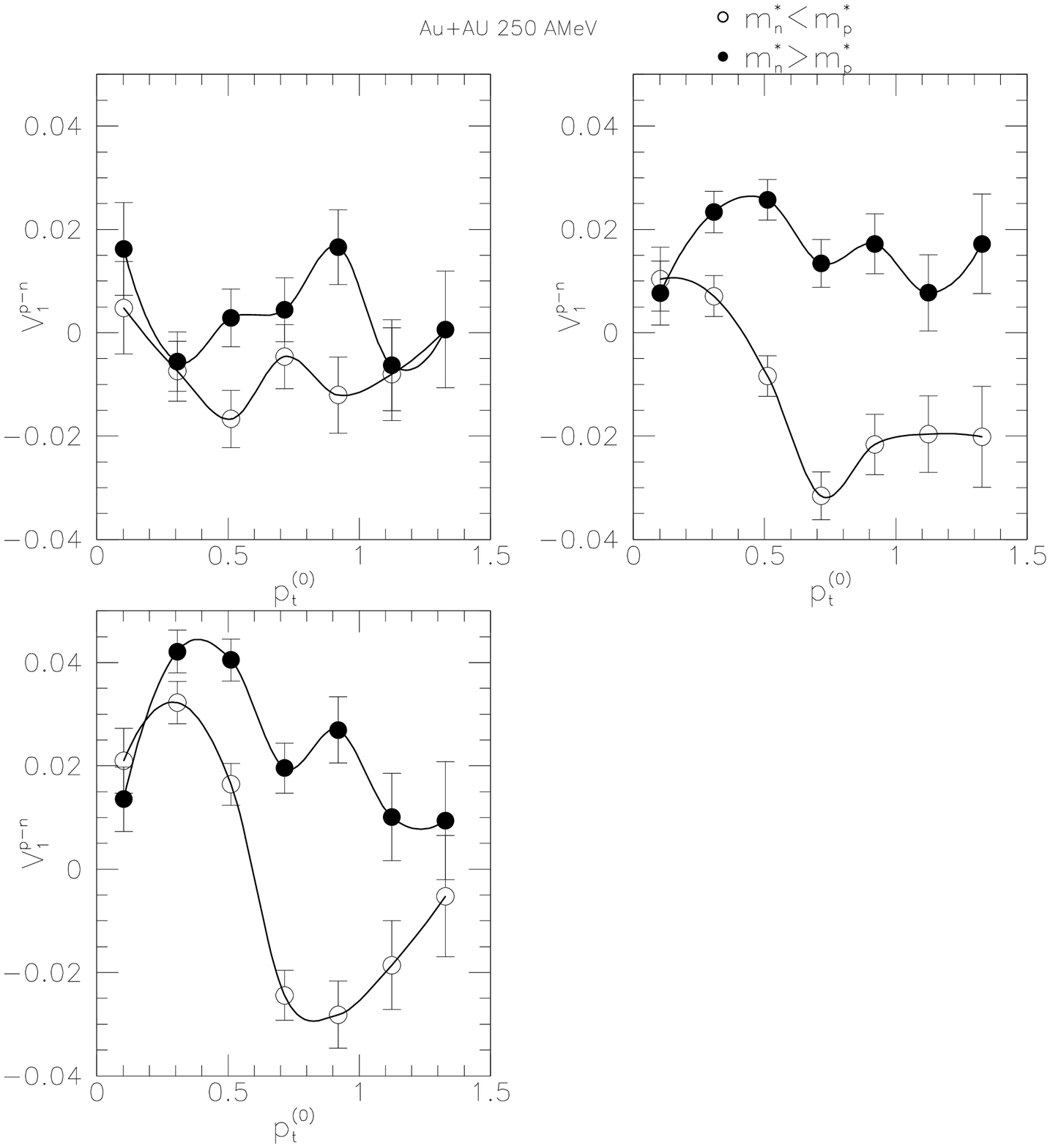}
\caption{\it Difference between proton and neutron $V_1$ flows in a 
semi-central
reaction Au+Au at 250\, AMeV for three rapidity ranges. Upper Left Panel:
$|y^{(0)}| \leq 0.3$; Upper Right: $0.3 \leq|y^{(0)}| \leq 0.7$;
Lower Left: $0.6 \leq|y^{(0)}| \leq 0.9$.
Lower Right Panel: Comparison of the $V_1$ proton flow 
with FOPI data \cite{fopi_v1} 
for three rapidity ranges. Top: $0.5 \leq|y^{(0)}| \leq 0.7$;
center: $0.7 \leq|y^{(0)}| \leq 0.9$; 
bottom: $0.9 \leq|y^{(0)}| \leq 1.1$.
} 
\label{v1dif}
\end{figure}

Collective flows are expected to be very sensitive to the momentum
dependence of the mean field, see \cite{BaranPR410} and refs.therein.
We have then tested the isovector part of the momentum dependence
just evaluating the $Differential$ transverse and elliptic flows
$$
V^{(n-p)}_{1,2} (y,p_t) \equiv V^n_{1,2}(y,p_t) - V^p_{1,2}(y,p_t)
$$ 
at various rapidities and transverse momenta in semicentral
($b/b_{max}=0.5$) $^{197}Au+^{197}Au$ collisons at $250AMeV$, where
some proton data are existing from the $FOPI$ collaboration at $GSI$
 \cite{fopi_v1,fopi_v2}.

\subsubsection{Transverse flows}
For the differential transverse flows, see Fig. \ref{v1dif} the mass splitting
effect is evident at all rapidities, and nicely increasing at larger
rapidities and transverse momenta, with more neutron flow when
$m_n^*<m_p^*$

Just to show that our simulations give realistic results we compare in 
lower right panel of Fig. \ref{v1dif} with the proton data of 
the $FOPI$ collaboration
for similar selections of impact parameters rapidities and transverse momenta.

The agreement is quite satisfactory. We see a slightly reduced proton flow
at high transverse momenta in the $m_n^*<m_p^*$ choice, but the effect
is too small to be seen fron the data. Our suggestion of the differential
flows looks much more promising.

\subsubsection{Elliptic flows}

\begin{figure}[htb]
\centering
\begin{picture}(0,0)
\put(135.0,2.6){\mbox{\includegraphics[width=4.5cm]{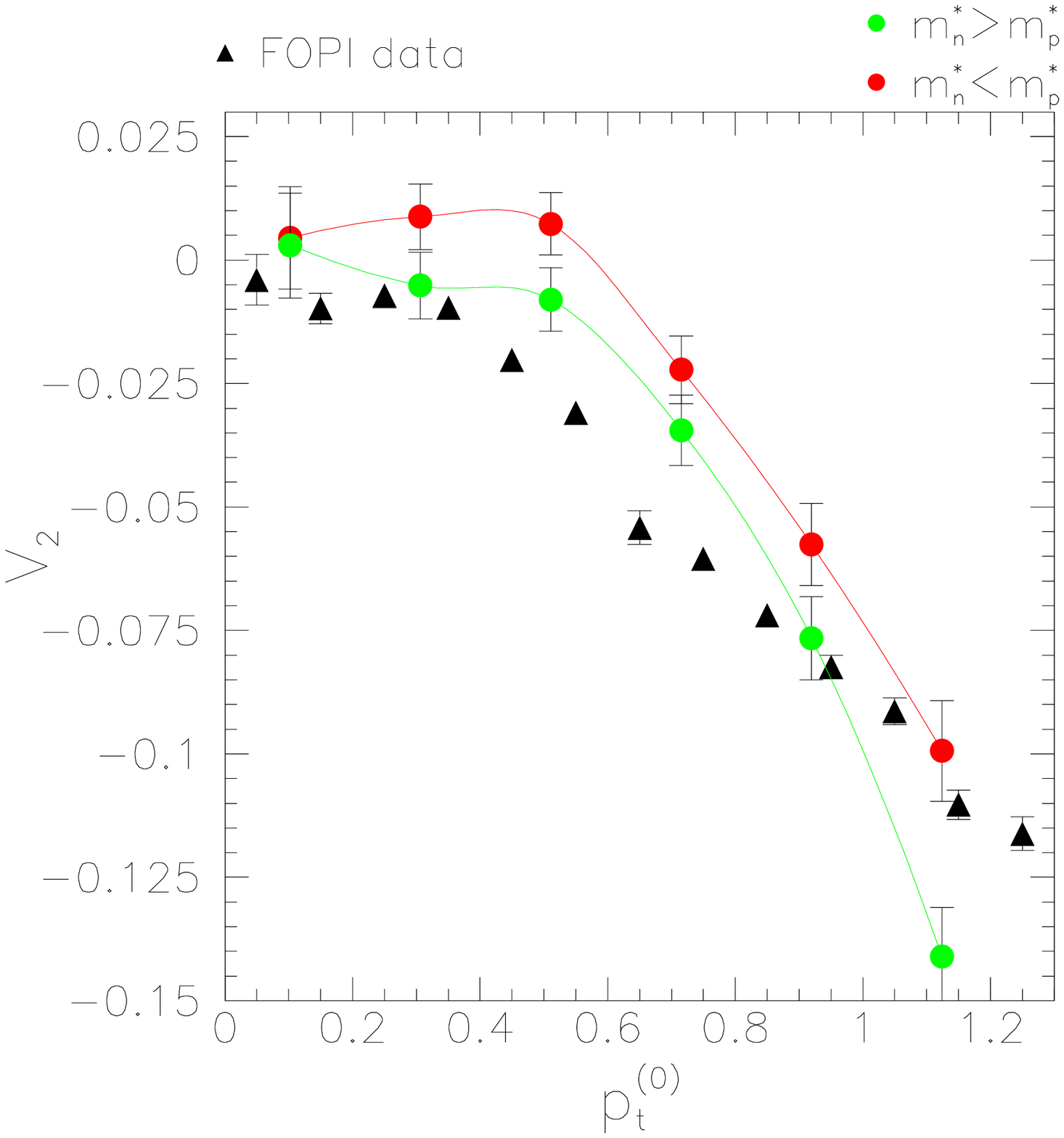}}}
\end{picture}
\includegraphics[width=10cm]{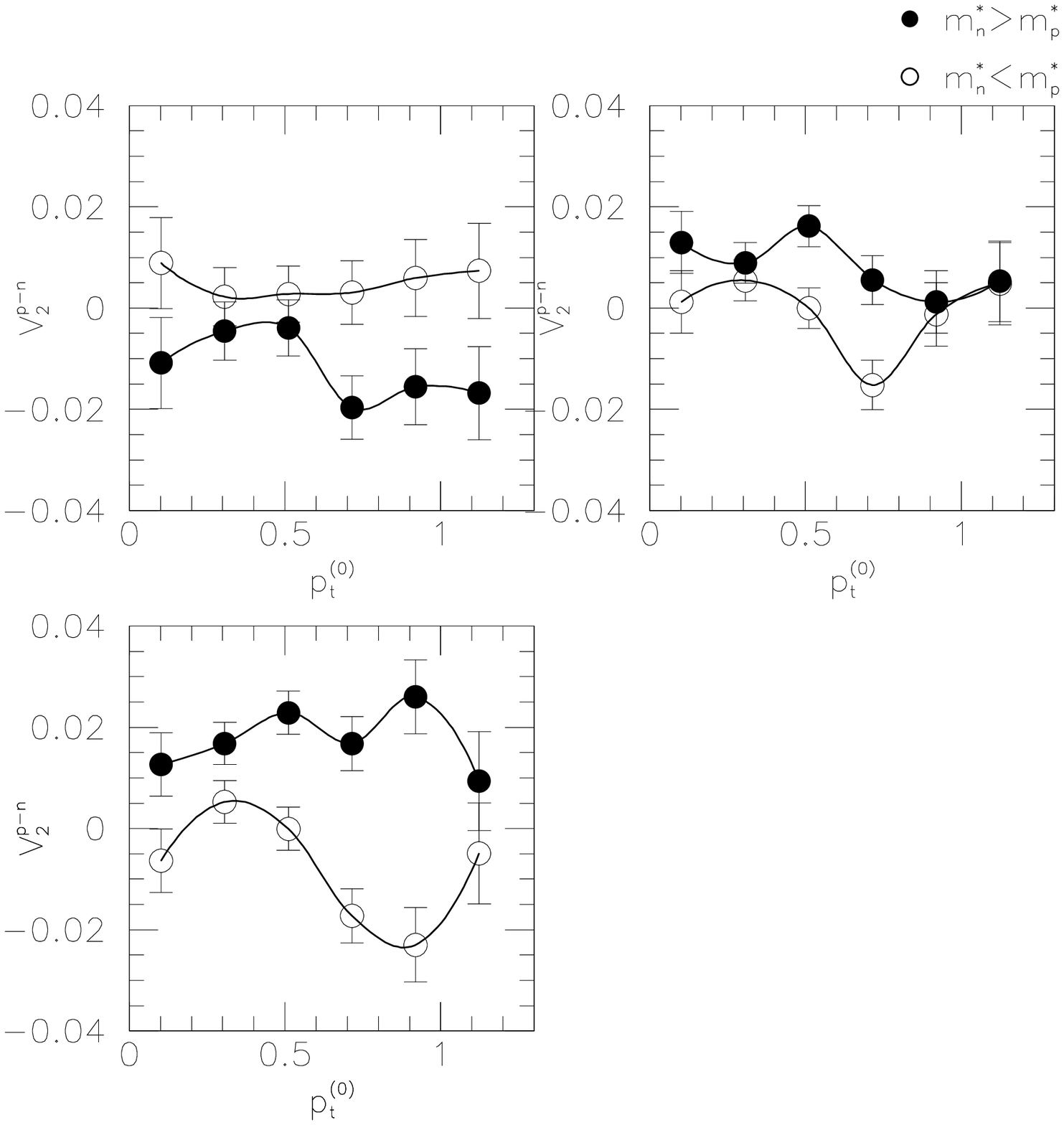}
\caption{\it Upper (left and right) and lower left panels: 
Difference between proton and neutron elliptic flows for the same
reaction and rapidity ranges as in Fig. \ref{v1dif}. Lower right panel:
Comparison of the elliptic proton flow with FOPI data
\cite{fopi_v2} (M3 centrality bin, $|y^{(0)}| \leq 0.1$).} 
\label{v2dif}
\end{figure}

The same analysis has been performed for the differential elliptic flows,
see Fig. \ref{v2dif}. Again the mass splitting effects are more evident
for higher rapidity and tranverse momentum selections. In particular
the differential elliptic flow becomes sistematically negative when
$m_n^*<m_p^*$, revealing a faster neutron emission and so more neutron
squeeze out (more spectator shadowing).

In the lower right panel we also show a
comparison with recent proton data from the $FOPI$ collaboration.
The agreement is still satisfactory. As expected the proton flow
is more negative (more proton squeeeze out) when $m_n^*>m_p^*$.
It is however difficult to draw definite conclusions only from proton data.

Again the measurement of differential flows appears essential.
This could be in fact an experimental problem due to the difficulties in
measuring neutrons. Our suggestion is to measure the difference between
light isobar flows, like triton vs. $^3He$ and so on.
We expect to clearly see the effective mass splitting effects, 
 maybe even enhanced due to larger overall flows, see 
 \cite{BaranPR410,ScalonePLB461}.


\begin{thebibliography}{00}
\bibitem{par} The $NL\rho$, $NL\rho\delta$ parametrizations used here have 
been derived 
from transport studies of relativistic heavy ion collisions, see Ch.8
of ref. \cite{BaranPR410}.
\bibitem{delta-corr} We like to remind that an isovector scalar channel
in the effective interaction ``automatically'' appear when correlations
are accounted for, see discussion in \cite{BaranPR410}.
\end{thebibliography}
\end{document}